\documentclass[prl,twocolumn,showpacs,superscriptaddress]{revtex4}
\usepackage{amsmath,amssymb,graphicx,epstopdf,color,bm,soul}

\begin{document}


\title{Phonon-Mediated Phase Transitions in Two-Dimensional Driven-Dissipative Systems}

\author{D. V. Karpov}
\affiliation{Institute of Photonics, University of Eastern Finland, P.O. Box 111 Joensuu, FI-80101 Finland}
\affiliation{ITMO University, St. Petersburg 197101, Russia}

\author{T. C. H. Liew}
\affiliation{Division of Physics and Applied Physics, Nanyang Technological University 637371, Singapore}

\author{I. G. Savenko}
\affiliation{A.V. Rzhanov Institute of Semiconductor Physics, Siberian Branch of Russian Academy of Sciences, Novosibirsk 630090, Russia}
\affiliation{Nonlinear Physics Centre, Research School of Physics and Engineering, The Australian National University, Canberra ACT 2601, Australia}

\begin{abstract}
We develop a two-dimensional stochastic dissipative theory for the description of the transport of exciton polaritons accounting for their interaction with the environment of acoustic phonons. Our approach is based on the explicit modeling of the corresponding microscopic processes using a Monte Carlo framework rather than modeling from phenomenological principles. We show the dynamic formation of a  condensate and investigate its characteristics, including threshold-like behavior in populations and the formation of spatial and temporal coherence at different temperatures of the environment and accounting for the stimulated nonlinear scattering, caused by system-environment interaction. The spatial coherence reveals a transition from an exponential to polynomial decay which can be attributed to the Berezinskii-Kosterlitzh-Thouless-like phase.
\end{abstract}

\maketitle


{\it Introduction.---} Exciton polaritons (EPs) in semiconductor microcavities represent intrinsically two-dimensional (2D) non-equilibrium systems that typically form a steady state where the total dissipation is balanced by the pumping performed over the system. Despite their nonequilibrium nature, several phase transitions have been studied in the system, the earliest example being that of polariton condensation~\cite{Kasprzak2006, Balili2007, Lai2007}, typically characterized by a spontaneous formation of coherence in the system. A polariton condensate usually differs from the conventional Bose-Einstein condensates (BEC), in particular in that it forms at the dynamic equilibrium rather than the thermal equilibrium, unless in long-lifetime samples~\cite{Balili2007}, and it may not necessarily appear in the lowest energy mode of the system~\cite{Maragkou2010, RefPRB93121303R2016}. Due to the nonequilibrium nature, the usual rules such as the non-existence of long-range order (Hohenberg-Mermin-Wagner theorem) require re-examination. Recent work revealed that no algebraic order can exist in the infinite non-equilibrium system, while order may exist up to a characteristic length scale (known as the Kardar-Parisi-Zhang length scale), which is anyway much larger than the size of typical exciton-polariton systems~\cite{Altman2015}.

The study of coherence of EPs is also relevant to research devoted to related phase transitions, including the transition to a superfluid state~\cite{ Wertz2010, AmoNature}, typically characterized by the absence of scattering and the maintenance of a \textit{frictionless flow}. In the mean time EP condensates were found to support topological defects (vortices)~\cite{Lagoudakis2008}, which, according to the equilibrium theory, are expected to disrupt long-range order and superfluidity in 2D systems at finite temperatures. These arguments motivated the study of interactions between vortices and antivortices in accordance with the Berezinskii-Kosterlitz-Thouless (BKT) transition~\cite{Fraser2009, Manni2011}. While the BKT transition is a well-defined concept in equilibrium, occurring when vortices and anti-vortices become bound together below some critical temperature, it was only until recently that numerical evidence arrived for an analogous transition in driven-dissipative systems, considering a coherently excited polariton optical parametric oscillator as an example~\cite{Dagvadorj2015}. It was pointed out that the corresponding theory for an incoherently driven polariton condensate is challenging due to the lack of a suitable model accounting for energy relaxation processes from first principles.

The superfluidity of EPs has been separately investigated. In Ref.~\onlinecite{KeelingPRL1070804022011} an analytical approach is developed, where using the Schwinger-Keldysh approaсh to nonequilibrium systems the author shows simultaneous co-existance of the normal and superfluid phases of the fluid at certain pump and decay. It should be noted that this decay is a crucial component which determines the normal fraction. At a phenomenological level, energy relaxation processes can be regarded as analogous to a change in the decay of different energy levels in the system: a high energy level experiences enhanced decay as it loses particles to a lower level, which effectively experiences reduced decay. Again, while it is anticipated that polariton superfluids would be robust against phonon-induced energy relaxation processes, it has not been possible to account for the possibility of such effects explicitly in theory, to our knowledge.

The theoretical modelling of the energy relaxation of EPs is a complicated task due to (i) the non-equilibrium nature of the particles and the presence of multiple channels of dissipation, (ii) the presence of partial energy relaxation and partial coherence, (iii) generally nonlinear behavior with both conservative and dissipative nonlinearities~\cite{ourPRB2015}. Even after the formation of the BEC in some actual single-particle ground or non-ground state, particles might endure scattering mediated by acoustic phonons or other bosons in the system to further relax or gain energy and change their phase~\cite{Tassone1997, Wouters2010}. By \textit{other bosons} we mean hot incoherent excitons, dark excitons and indirect excitons, however, we leave the theoretical study of these dissipation channels for future work.


There have been suggested several approaches for treating the energy relaxation of particles. Some of them operate with white noise and spontaneous scattering~\cite{Carusotto2005, PRB771153402008, PRB791653022009} and thus stimulated scattering towards the ground state cannot be accounted for. Other approaches account for the correlations in the system~\cite{Savenko2011}, however, it seems impossible to describe systems of dimensionality higher than one~\cite{PRLCoherence, JETP}. In our previous work~\cite{RefPRL1101274022013} we started to develop a microscopic theoretical model which accounts for the effects of partial energy relaxation of EPs via the scattering with the acoustic phonon field, thus suggesting an alternative to phenomenological white noise based approaches. Now we consider a 2D EP system, which is the more common arrangement studied experimentally and we show that energy relaxation plays there a dramatic role.

\begin{figure}[!t]
	\hspace{10pt}
	\includegraphics[width=0.9\linewidth]{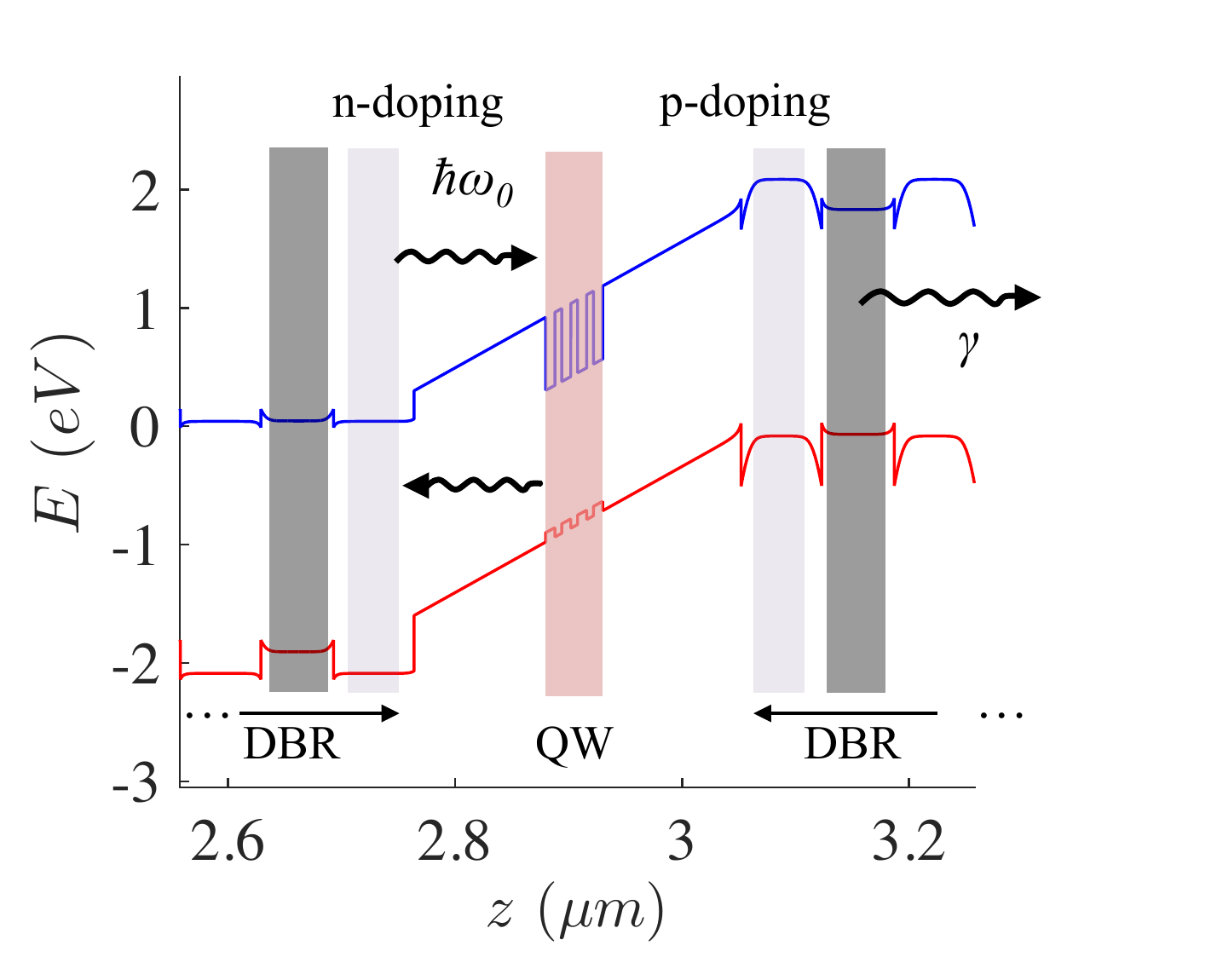}
	\caption{(color online) A typical InAlGaAs heterostructure-based single-mode ($\hbar\omega_0$) microcavity consisting of four quantum wells (QWs) located between two distributed Bragg mirrors (DBRs) with radiative losses $\gamma$. Blue and red curves show the Conduction and Valence bands calculated using~\cite{SiLENSe}, respectively.}
	 \label{Fig1}
\end{figure}

{\it Theoretical model.---} Let us consider a microcavity presented in Fig.~\ref{Fig1} in $\hat z$ (growth) direction.
We model the 2D transport of EPs in the $xy$ plane using the approach or Refs.~\cite{RefPRL1101274022013, ourPRB2015}, operating with the time-dependent EP field operator, $\hat{\Psi}(\mathbf{r},t)$, which, within the mean field approximation is replaced by a classical variable, $\psi(\mathbf{r},t)=\langle\hat{\Psi}(\mathbf{r},t)\rangle$. The Fourier image of the EP field is $\psi_{\mathbf{k}_\parallel}(t)={\cal F}[\psi(\mathbf{r},t)]=(1/N)\sum_{\mathbf{r}}\psi(\mathbf{r},t)e^{-i{\mathbf{k}_\parallel}\cdot\mathbf{r}}$, where $N$ is the discretization number, ${\mathbf{r}}$ is a 2D coordinate and ${\mathbf{k}_\parallel}$ is the in-plane wave vector:
\begin{eqnarray}
\label{EqGPE}
i\hbar\frac{d\psi({\mathbf{r},t})}{dt}={\cal F}^{-1}\left[E_{\mathbf{k}_\parallel}\psi_{\mathbf{k}_\parallel}(t)+{\cal S}_{\mathbf{k}_\parallel}(t)\right]\\
\nonumber
+\left[i\frac{\hbar }{2}Rn_X-i\frac{\hbar\gamma}{2}+\alpha_N\left|\psi({\mathbf{r}},t)\right|^2
\right]
\psi({\mathbf{r}},t)\\
\nonumber
+\sum_{\mathbf{k}_\parallel}\left\{{\cal T}_{-{\mathbf{k}_\parallel}}(t)+{\cal T}^*_{\mathbf{k}_\parallel}(t)\right\}\mathrm{e}^{-i{\mathbf{k}_\parallel}\cdot{\mathbf{r}}}
\psi({\mathbf{r}},t).
\end{eqnarray}
Here the first term $E_{\mathbf{k}_\parallel}$ is the bare dispersion of bosons, which is non-parabolic for EPs;
$n_X$ is the occupation of an incoherent exciton reservoir, which dynamics we discuss below;
$R$ is the reservoir-system excitations exchange rate; and the term $-i\hbar\gamma\psi({\mathbf{r}},t)/2$ accounts for the finite radiative lifetime of the particles~\cite{Carusotto2004}. $\alpha_N$ is the strength of nonlinear EP interaction, which can be estimated as~\cite{Yamamoto1999}: $\alpha_N\approx E_ba_B^2/(\Delta x\Delta y)$, where $\Delta x=L_x/N$, $\Delta y=L_y/N$ are the discretization units, with $L_x\times L_y$ being the cavity 2D area.

The interaction with acoustic vibrations of the crystal lattice (referred to as phonons in the following) leads to the appearance of two types of terms.
The first term is the particle number stimulated phonon-mediated scattering term:
\begin{equation}
{\cal S}_{\mathbf{k}_\parallel}(t)=\sum_{\mathbf{q_\parallel}}\psi_{{\mathbf{k}_\parallel}+{\mathbf{q}_\parallel}}(t)\left(\int_0^t{\cal A}_{{\mathbf{q}_\parallel}}(t^\prime){\cal K}_{0}(t-t^\prime)dt^\prime\right),
\label{EqStimulated}
\end{equation}
where ${\cal A}_{{\mathbf{q}_\parallel}}(t)=\sum_{\mathbf{k}_\parallel^\prime}\hat{a}_{{\mathbf{k}_\parallel}^\prime+{\mathbf{q}_\parallel}}^\dagger\left(t\right)\hat{a}_{{\mathbf{k}_\parallel}^\prime}\left(t\right)$.
It should be noted that the convolution integral in Eq.~\eqref{EqStimulated} takes care of the energy conservation.
Here we use the phonon wave vector, $\mathbf{q}=\hat{e}_xq_x+\hat{e}_yq_y+\hat{e}_zq_z$, where $\hat{e}_x$, $\hat{e}_y$ and $\hat{e}_z$ are unit vectors along corresponding axes: $\hat{e}_x$, $\hat{e}_y$ lie in the in-plane of the cavity, thus $\mathbf{q}_\parallel=q_x\hat{e}_x+q_y\hat{e}_y$, whereas $\hat{e}_z$ is in the structure growth direction.
The dispersion of the phonons reads $\hbar\omega_{\mathbf{q}}=\hbar
u_s\sqrt{q_x^2+q_y^2+q_z^2}$. It is determined by the sound velocity, $u_s$.
In Eq.~\eqref{EqStimulated},
\begin{eqnarray}
\nonumber
{\cal K}_{{\mathbf{q}_\parallel}}(t)= i\frac{L_z}{\pi}\int_{-\pi/L_z}^{\pi/L_z}|G({\mathbf{q}_\parallel})|^2\sin[\omega({\mathbf{q}_\parallel})t]dq_z
\end{eqnarray}
turns out approximately independent of $|{\mathbf{q}_\parallel}|$ in the range of
$|{\mathbf{q}_\parallel}|\in(-10^8,10^8)$ $m^{-1}$, and thus in our calculations we put
${\cal K}_{{\mathbf{q}_\parallel}}(t)\approx{\cal K}_0(t)$.
Here $G({\mathbf{q}_\parallel})$ is the exciton-phonon interaction strength, whose
calculation can be found elsewhere~\cite{Piermarocchi1996, Carmichael, Hartwell2010}.

The stochastic functions ${\cal T}_{{\mathbf{q}_\parallel}}$ in the last line of Eq.~\eqref{EqGPE} read
\begin{align}
\left<{\cal T}_{\mathbf{q}_\parallel}^*(t){\cal T}_{\mathbf{q}^\prime_\parallel}(t^\prime)\right>&=
\sum_{q_z}\left|G_{{{\mathbf{q}_\parallel},q_z}}\right|^2n_{{\mathbf{q}_\parallel},q_z}\delta_{{\mathbf{q}_\parallel},{\mathbf{q}^\prime_\parallel}}\delta(t-t^\prime),\notag\\
\left<{\cal T}_{{\mathbf{q}_\parallel}}(t){\cal T}_{{\mathbf{q}^\prime_\parallel}}(t^\prime)\right>&=
\left<{\cal T}_{{\mathbf{q}_\parallel}}^*(t){\cal T}_{{\mathbf{q}^\prime_\parallel}}^*(t^\prime)\right>=0.
\label{EqThermal}
\end{align}
These terms contain the phonon density, $n_{{\mathbf{q}_\parallel},q_z}$, and thus they are temperature dependent.
We calculate $\psi({\mathbf{r},t})$ over multiple realizations of the evolution of the system with stochastic variables ${\cal T}_{\mathbf{q}_\parallel}(t)$. The last term in~\eqref{EqGPE} contains the first power of $\psi({\mathbf{r},t})$, which means that the stochastic phonon-mediated scattering acts as an effective nonradiative lifetime, corresponding to the absorption and emission of the phonons by the EP ensemble. $n_X(\mathbf{r},t)$ describes the density of reservoir excitons and evolves according to the rate equation:
\begin{eqnarray}
\label{EqExcitons}
\frac{\partial n_X(\mathbf{r},t)}{\partial t}=P-\frac{n_X}{\tau_X}-R~ n_X|\psi(\mathbf{r},t)|^2.
\end{eqnarray}
Here $\tau_X$ is the exciton lifetime, $P$ is the incoherent pumping power, and $\gamma$ is the rate of EP formation fed by the excitonic reservoir.
In the case of nonresonant electrical pumping of the system, $P$ can be expressed via electron and hole concentrations using the drift-diffusion model~\cite{ourAPL2016}: $P=W~n~p$, where $n$ and $p$ are electron an hole concentrations, calculated from the electron and hole Fermi quasi-energies, $W$ is a parameter (see also Fig.~\ref{Fig1} and~\cite{SiLENSe}).

{\it Threshold of condensation.---} In computations we use the parameters typical for InGaAlAs alloys: speed of
sound $u=5370$ $m/s$~\cite{Hartwell2010}, $\gamma=1/18$
$ps^{-1}$~\cite{Gao2012}.
\begin{figure}[!t]
	\includegraphics[width=0.99\linewidth]{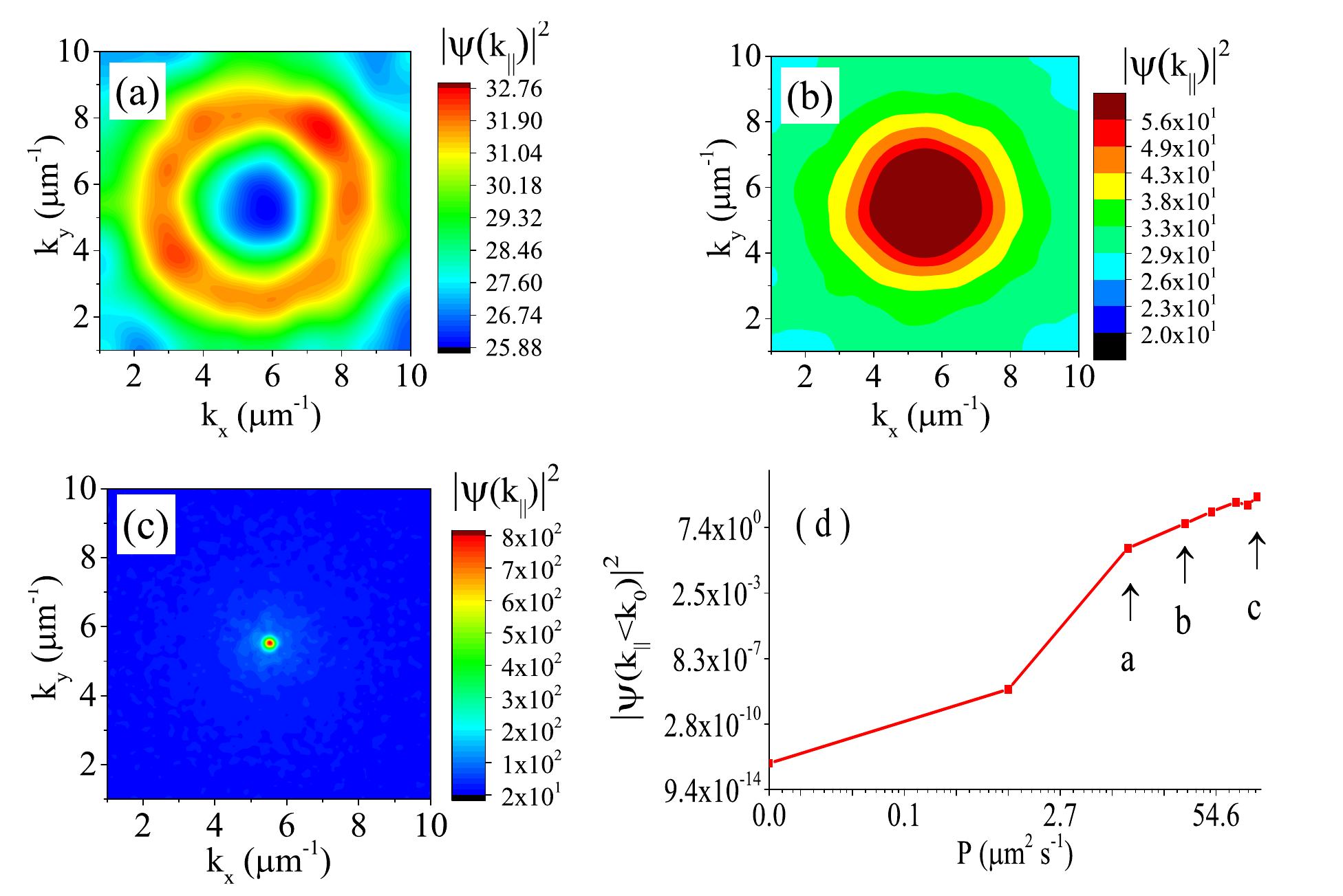}
	\caption{(color online) Exciton-polariton distribution in momentum space for different pumping powers: (a) $P=10~\mu$m$^{2}$s$^{-1}$, nearly-equilibrium distribution, far below condensation threshold, the bottleneck effect; (b) $P=50~\mu$m$^{2}$s$^{-1}$, close to the condensation threshold, bottleneck effect takes place; (c) $P=100~\mu$m$^{2}$s$^{-1}$, above threshold, EPs condense in the single-particle ground state. (d) shows the threshold characteristics in the log scale.}
		\label{Fig2}
\end{figure}
The EP dispersion was calculated using a two oscillator model with cavity photon effective mass $4\times10^{-5}$ of the free electron mass, Rabi splitting $10$ meV and exciton-photon detuning $2.5$ meV at zero in-plane wave vector.

\begin{figure}[!b]
	\includegraphics[width=0.95\linewidth]{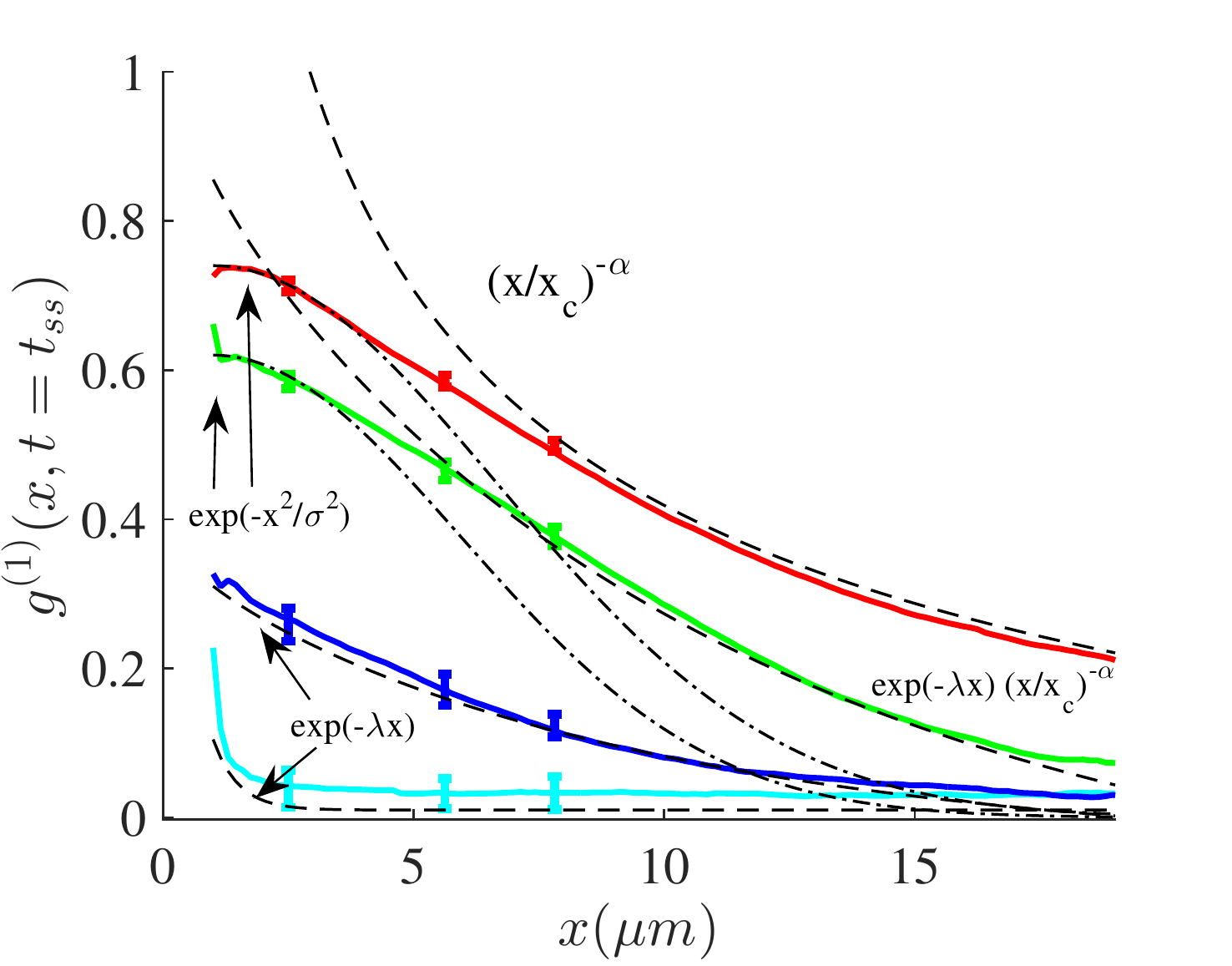}
	\caption{(color online) First-order correlation function at $T=10$ K in the steady state, $t_{ss}=200~$ps, for different pumping strengths, from bottom to top: cyan, blue, green, and red solid curves correspond to pumps $P = 1,~10,~50,~100~ \mu$m$^{2}$ s$^{-1}$, respectively. Two lower solid curves are approximated by exponents; Two upper curves are approximated by Gaussian at small $x<5$ $\mu$m.  At $x>5$ $\mu$m, the third from the bottom (green) solid curve is approximated by the general power law manifesting the transition from exponential to polynomial decay. The upper (red) solid curve is well approximated by the power-law decay starting at $x\approx 7$ $\mu$m.}
	\label{Fig3}
\end{figure}

\begin{figure}[!t]
	\includegraphics[width=0.79\linewidth]{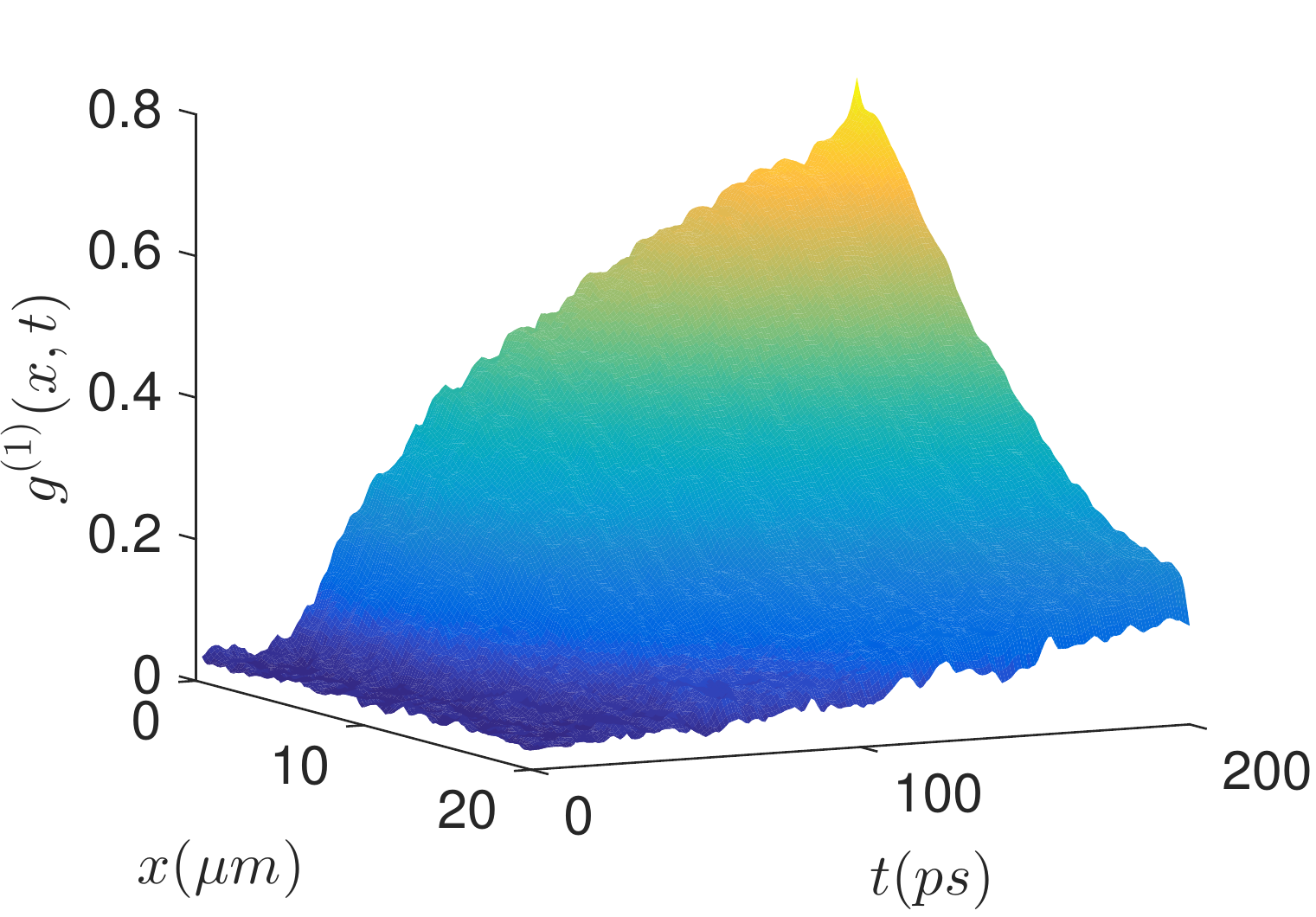}
	\caption{(color online) Evolution of the first-order correlation function at $T=10$ K at $P = 100$ $\mu$m$^{2}$s$^{-1}$ in space and time: manifestation of the spatial coherence establishment in the system.}
	\label{Fig4}
\end{figure}

\begin{figure}[!b]
	\includegraphics[width=0.9\linewidth]{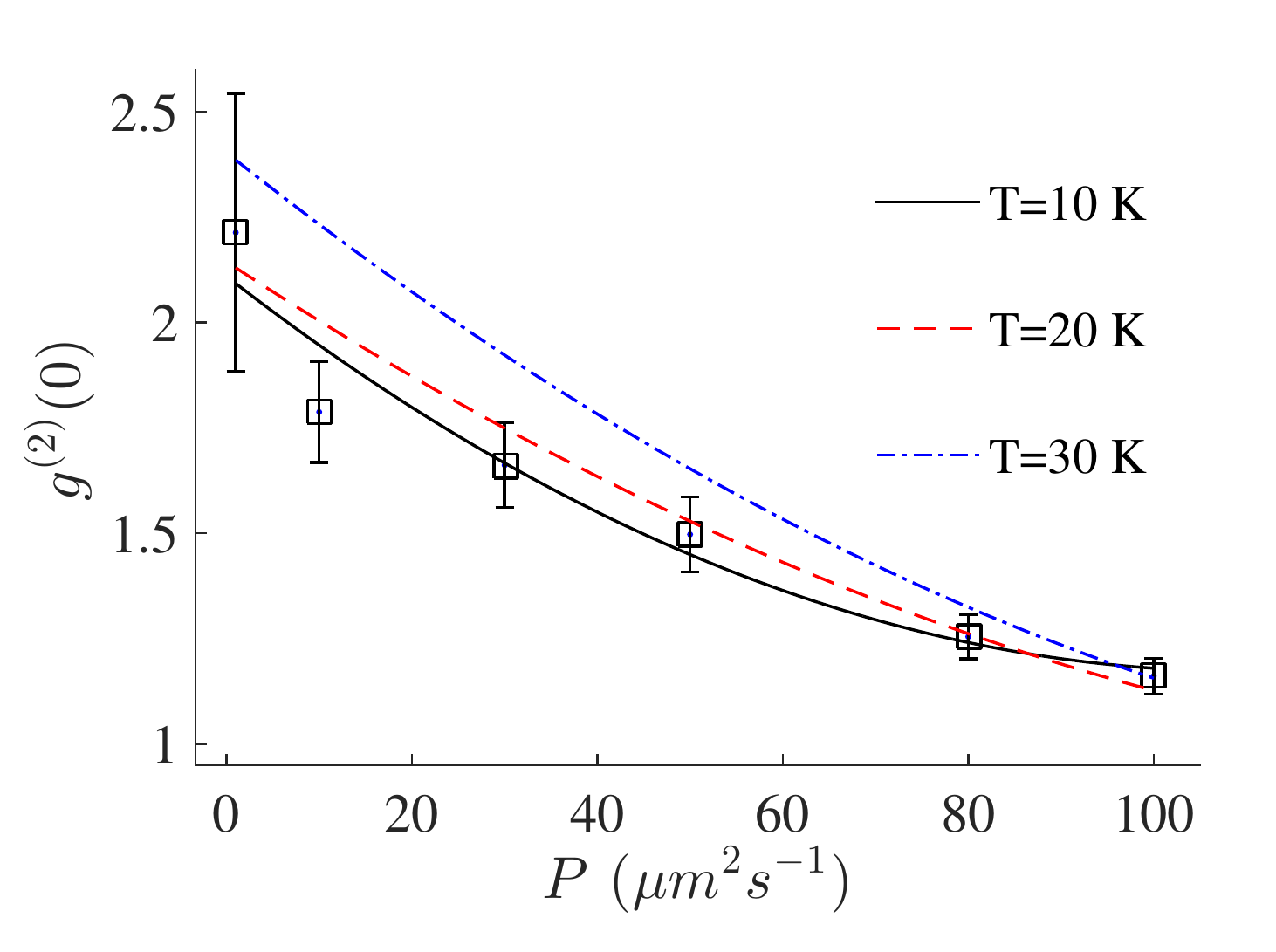}
	\caption{(color online) Zero-delay second-order temporal coherence function: the second-order correlator as a function of pumping power for different temperatures. Black, red and blue curves correspond to $T =10,~20,~30~K$, respectively. Error bars are shown for the black curve only. }
	\label{Fig6}
\end{figure}

Solving numerically coupled equations~\eqref{EqGPE} and~\eqref{EqExcitons} and averaging over stochastic trajectories, we obtain the macroscopic EP wave function (and the correlation functions described below in Eqs.~\eqref{EqSpatialCoh} and~\eqref{EqTemporalCoh}). Stochastic variables are introduced as a normalized random noise depending on temperature. The number of trajectories is chosen equal to 400 (which is enough to achieve convergence of the results with an appropriate accuracy; see also error bars in Figs.~\ref{Fig3}-\ref{Fig6}).

Figure~\ref{Fig2} shows the formation of the BEC.
Panel (a) corresponds to the bottleneck state observed experimentally~\cite{Richard2005}, when the particles are locked at the higher energy states near the dispersion inflection point. The bottleneck is attributed to inefficiency of phonon-induced energy relaxation of EPs in the region of their dispersion that is significantly steeper than the phonon dispersion when nonlinear particle-particle interaction and stimulated phonon scattering are weak at the corresponding pumping power.
Panel (b) corresponds to still thermal state of the system, when EPs are close to the condensation threshold.
Panel (c) shows the particle distribution above threshold. The increase of the EP density allows stimulated scattering processes to overcome the bottleneck region and form a condensate~\cite{Kasprzak2006} characterized by the collection of a macroscopic population in the ground state (and the onset of coherence, as measured by the second order correlation function below).
Panel (d) shows the occupation of the ground state as a function of pumping power. Clear threshold-like characteristic is visible, followed by a rapid increase of EP density in the ground state, corresponding to the formation of the EP BEC.
	
{\it Spatial and temporal coherence.---} We further investigate the coherence properties of the system.
The first-order spatial coherence function, by definition, reads:
\begin{align}
g^{(1)}(\mathbf{r}_{\parallel},t)&=\frac{\langle\psi^{*}(0,t_{ss})\psi(\mathbf{r}_{\parallel},t) \rangle}{\sqrt{\langle|\psi(0,t_{ss})|^2\rangle \langle|\psi(\mathbf{r}_{\parallel},t)|^2\rangle}},
\label{EqSpatialCoh}
\end{align}
where the ensemble averaging takes place at the steady state, $t=t_{ss}$. Since we are considering homogeneous pumping, $\mathbf{r}_{\parallel}$ can be replaced by ${r}_{\parallel}=|\mathbf{r}_{\parallel}|$. It should be noted, however, that one can also use inhomogeneous in $x$-space pumping or inhomogeneous structure, such as a lattice, and analyze coherence at any spot, $\mathbf{r}_{\parallel}$, within the approach.

Figure~\ref{Fig3} shows the first-order correlation function for $T=10$ K in the steady state, $t_{ss}\approx 200$ ps and later, as a function of ${r}_{\parallel}$ in the plane of the QW for various pumping strengths. 
We observe a behavior which is typical for a BKT transition. Indeed, at weak pumps, $P = 1,~10~\mu$m$^{2}$ s$^{-1}$, the decay is exponential, $\sim\exp(-\lambda x)$ which corresponds to the absence of any long-range order in the system. At $P=50~\mu$m$^{2}$ s$^{-1}$ (green solid curve), we come to the transition point, when the coherence function is well approximated by the Gaussian function, $\exp(-x^2/\sigma^2)$, at small $x$ and, more importantly, the general power law, $\sim\exp(-\lambda x)\cdot(x/x_c)^{-\alpha}$ at $x>5$ $\mu$m. Here $\sigma=7$ $\mu$m, $\lambda=0.08(3)$ $\mu$m$^{-1}$, $x_c=6$ $\mu$m and $\alpha=0.05$. 
It should be noted, that the two lower solid curves (cyan and blue) can certainly be also fitted with the general power law, however, on both of these curves $\alpha\rightarrow 0$.

The upper (red) solid curve can also be approximated with $\exp(-\lambda x)\cdot(x/x_c)^{-\alpha}$ after $\sim 7$ $\mu$m, however, here $\lambda\rightarrow 0$, whereas $\alpha\approx0.45$ (compare with the green curve). Other parameters are: $\sigma=8$ $\mu$m, $x_c=6$ $\mu$m for the upper (red) curve. Therefore it is fair to say, that the upper curve decays as a polynomial $(x/x_c)^{-0.45}$.
%
%
%
%
%
%
In addition, in Fig.~\ref{Fig4} we show the evolution and establishment of the first-order coherence in the system. The function $g^{(1)}$ increases with time (at high enough pump $P = 100$ $\mu$m$^{2}$s$^{-1}$) and it vanishes at certain distances as a power law (See upper (red) curve in Fig.~\ref{Fig3}). 
An interesting feature of the power law is that it is impossible to attribute a characteristic distance to it defining it as coherence length $r_c$ (as opposed to the exponential decay, where $g^{(1)}(r_c)=e^{-1}$).
%
%
%
%
%
%

Further, we define the second-order temporal coherence function as
\begin{align}
g^{(2)}(t)&=\frac{\langle\psi^{*}(\mathbf{r}_{\parallel},t_{ss})\psi^{*}(\mathbf{r}_{\parallel},t)\psi(\mathbf{r}_{\parallel},t_{ss})\psi(\mathbf{r}_{\parallel},t)\rangle}{\langle|\psi(\mathbf{r}_{\parallel},t_{ss})|^2\rangle \langle|\psi(\mathbf{r}_{\parallel},t)|^2\rangle},
\label{EqTemporalCoh}
\end{align}
and it also depends on the pumping power and temperature. Here we also consider homogeneous pumping. 

Figure~\ref{Fig6} shows the second-order correlation function for zero delay ($t-t_{ss}=0$) for different temperatures as a function of pumping power.
For low pumping powers, the statistics is thermal and the mean value of the correlator approximately equals two. however, with the increase of power it drops down that corresponds to the formation of macroscopic coherence in the system (formation of BEC), see also Fig.~\ref{Fig2}.
With the increase of temperature, the correlations between different states become weaker, as expected.




{\it Conclusions.---} We have developed a two-dimensional stochastic dissipative Gross-Pitaevskii
equation, where the energy relaxation of the system is mediated by a field of incoherent bosons represented by, for instance, acoustic phonons, as in our case. We have investigated the dynamics of the system and calculated general coherence properties, in particular the spatial and temporal coherence functions. Further we have suggested a way to describe the BKT phase transition for exciton polaritons at finite temperatures.

In order to assay our approach, we have applied it first to model existing effects and showed that it consistently reproduces a number of features of modern experiments, in particular, the bottleneck effect~\cite{Richard2005}, in which EPs become trapped in a region of high gradient in their dispersion. Further, we investigate the behavior of spatial and temporal coherence functions under homogeneous excitation. The first order coherence function is found to exhibit a Gaussian decay at small distances and a power law decay at longer distances at certain excitation conditions. This is consistent with predictions based on the BKT transition~\cite{PRB2014}. When analyzing the second order coherence, we observe a gradual transition from a thermal state towards a coherent statistics with increasing pump power. Here, we find that it is very challenging to obtain a perfectly coherent state, which is in agreement with previous theoretical~\cite{Schwendimann2008} and experimental~\cite{Love2008} reports. Indeed, as it has been pointed out~\cite{Kim2015}, the presence of multiple modes lying nearby in energy prevents the establishment of full coherence unless geometries with discrete densities of states are employed.

We thank H. Flayac, M. Sun, O. Egorov, Yu. Rubo for useful and encouraging discussions.
We acknowledge support of the Australian Research Council Discovery Projects funding scheme (Project No. DE160100167); President of Russian Federation (Project No. MK-5903.2016.2). 
IGS thanks the support from the RNF (project RSF 17-12-01039).


\end{document}